\documentclass[fleqn,twoside]{article}
\usepackage{espcrc2}

\usepackage{graphicx}
\usepackage[figuresright]{rotating}

\newcommand{\AmS}{{\protect\the\textfont2
  A\kern-.1667em\lower.5ex\hbox{M}\kern-.125emS}}
\newcommand{\be}{\begin{equation}}
\newcommand{\ee}{\end{equation}}
\newcommand{\Real}{\mbox{Re}}
\newcommand{\Tr}{\mbox{Tr}}
\newcommand{\bea}{\begin{eqnarray}}
\newcommand{\eea}{\end{eqnarray}}

\newcommand{\nn}{\nonumber}

\newcommand{\plaquette}{\begin{array}{l}\begin{picture}(20,25)

        \put (0,0) {\vector(1,0){13}}
        \put (0,0) {\line(1,0){20}}
        \put (20,0) {\vector(0,1){13}}
        \put (20,0) {\line(0,1){20}}
        \put (20,20) {\vector(-1,0){13}}
        \put (20,20) {\line(-1,0){20}}
        \put (0,20) {\vector(0,-1){13}}
        \put (0,20) {\line(0,-1){20}}

\end{picture}\end{array}}
\newcommand{\rectangle}{\begin{array}{l}\begin{picture}(40,25)

        \put (0,0) {\line(1,0){20}}
        \put (0,0) {\vector(1,0){13}}
        \put (20,0) {\line(1,0){20}}
        \put (20,0) {\vector(1,0){13}}  
        \put (40,0) {\vector(0,1){13}}
        \put (40,0) {\line(0,1){20}}
        \put (40,20) {\vector(-1,0){13}}
        \put (40,20) {\line(-1,0){20}}
        \put (20,20) {\vector(-1,0){13}}
        \put (20,20) {\line(-1,0){20}}
        \put (0,20) {\vector(0,-1){13}}
        \put (0,20) {\line(0,-1){20}}

\end{picture}\end{array}}

\title{Improved Hamiltonian Lattice Gauge Theory}

\author{J. Carlsson\address[unimelb]{School of Physics, 
        University of Melbourne, \\ 
        Victoria 3010,
        Australia.}\thanks{\texttt{j.carlsson@physics.unimelb.edu.au}},
        J. A. L. McIntosh\addressmark\thanks{Presenter at the conference.},
        B. H. J. McKellar\addressmark and
        L. C. L. Hollenberg\addressmark}
       
\begin{document}

\begin{abstract}
We derive an improved lattice Hamiltonian for pure gauge
theory, coupling arbitrarily distant links 
in the kinetic term. The level of improvement achieved is examined in
variational calculations of the SU(2) specific heat in 2+1 dimensions.   
\vspace{1pc}
\end{abstract}

\maketitle

\section{Introduction}

To date, the majority of work in lattice QCD has been
performed in the action formulation. An advantage of this
approach is that it readily lends itself to Monte Carlo techniques.
Working in the Hamiltonian approach~\cite{Kogut:1975ag} brings a different
intuition to the problem and serves as a check of universality. An
advantage of Hamiltonian lattice gauge theory is in the applicability of 
 techniques from
many body physics~\cite{McKellar:2000zk}. Also, it appears that in finite
density QCD, a Hamiltonian approach is favorable due to the so-called
complex action problem which rules out the use of standard 
Monte Carlo techniques in the action
formulation~\cite{Gregory:2000pm}.

Much work in the past decade has been devoted to improving lattice
actions~\cite{Lepage:1996jw}.  
In contrast, the improvement of lattice
Hamiltonians has only recently begun. Perhaps the most extensive treatment to
date is due to Luo, Guo, Kr\"oger and Sch\"utte~\cite{Luo:1999dx} 
who discussed the improvement
of Hamiltonian lattice gauge theory for gluons.
In their study it was discovered that deriving an
improved Hamiltonian from a Symanzik improved action, whether by transfer
matrix or canonical Legendre transformation, results in a kinetic Hamiltonian
with an infinite number of terms coupling lattice sites which are
arbitrarily far apart. To derive a local kinetic\textsc{}
Hamiltonian coupling only nearest neighbor lattice sites it was found
necessary to start with an improved action with an infinite number
of terms, coupling distant lattice sites.

With this technique the order $a^2$ errors are removed from the
Kogut-Susskind Hamiltonian.  However, generating Hamiltonians with
further improvement would seem exceedingly difficult.  This is because
one would need to start from a L\"uscher-Weisz improved action with
non-planar terms~\cite{Luscher:1985zq}.  For this reason we propose a
move to the Symanzik approach, as applied to the Hamiltonian.  That
is, in the spirit of the original Kogut-Susskind paper, to construct
improved Hamiltonians directly by adding appropriate gauge invariant
terms and fixing their coefficients so that errors are canceled. 

\section{Symanzik Improvement of the Lattice Hamiltonian}

The Kogut-Susskind Hamiltonian for pure SU($N$) gauge theory on the
lattice is given by
\bea
H_0 = \frac{a^3}{2} \sum_{x,i} \Tr \left[E^L_i(x)^2
\right] + \frac{2N}{ag^2}\sum_{x, i<j} P_{ij}(x). 
\eea
Here $E^L$ is the lattice chromo-electric field, $N$ is the dimension
of the gauge group, and $P_{ij}(x)$ is the usual
plaquette and operator in the $(i,j)$ plane.

To improve the potential part of the Kogut-Susskind Hamiltonian, we follow
the process of improving the Wilson action. By introducing the usual rectangle
operator  $R_{ij}(x)$ in the $(i,j)$ plane (with 
the long side in the $i$ direction), and expanding in powers of $a$, 
we arrive at the order $a^2$ improved potential term:
\be
V_2 \!= \!\frac{2N}{ag^2}\!\sum_{x,i<j}\!\left\{\frac{5}{3} P_{ij}(x)\! -\!
\frac{1}{12}\left[R_{ij}(x)\!+\!R_{ji}(x)\right]\right\}.\label{vimp1}
\ee
To improve the kinetic part of the Kogut Susskind Hamiltonian we
proceed as follows. The {\em lattice} gluon field $A^L_\mu$ is defined
to be the average of the
continuum gluon field $A$ along the link joining $x$ and $x+a\mu$:
\be
A^L_\mu(x') = \frac{1}{a} \int_{\mbox{\tiny Link}} \hspace{-0.3cm}dx\cdot A
\,\Rightarrow \, U_\mu(x) = e^{iga A^L_\mu(x')}, \label{latticegf}
\ee 
where $x'$ is a point near the points $x$ and $x+a\mu$.
On the lattice, the gluon field is defined at only one
point along (or nearby) a link. This leads to interpolation errors in
in Eq.~(\ref{latticegf}). For instance, by choosing to
evaluate the gluon field at the midpoint of the link,
the lattice and continuum fields are related by
\be
A^L_\mu(x) \approx A_\mu(x) + \frac{a^2}{24}\partial_\mu^2 A_\mu(x) + 
\frac{a^4}{1920}\partial_\mu^4 A_\mu(x).
\ee
Since the electric field must generate group transformations,
$[E^\alpha_i(x), A^\beta_j(y)] = -i\delta_{xy}
\delta_{ij}\delta_{\alpha\beta}/a^3$, a similar relation between the
lattice and continuum electric fields can be derived:
\be
E^{L\alpha}_i(x)\approx E^\alpha_i(x) - \frac{a^2}{24}\partial_i^2
E^\alpha_i(x) 
+ \frac{7a^4}{5760}\partial_i^4 E^\alpha_i(x).\label{e2}
\ee

Making use of this approximation, the   
classical errors arising in the kinetic Hamiltonian can be determined. 
To cancel these errors we take the approach of adding new terms and
fixing their coefficients in order to cancel the order $a^2$ error. 
There is a great deal of freedom in choosing additional terms, which are
restricted only by gauge invariance and the need for an appropriate continuum 
limit.

To understand the construction of gauge invariant kinetic terms,
recall that the electric field and link operator
transform as follows under a local gauge transformation $\Lambda(x)$:  
\bea
E_i(x) &\rightarrow & \Lambda(x) E_i(x) \Lambda^\dagger(x) \\
U_i(x) &\rightarrow & \Lambda(x) U_i(x) \Lambda^\dagger(x+ai).
\eea
Consequently, the next most
complicated gauge invariant term we can construct (after $\Tr E^LE^L$)
couples nearest neighbor electric fields, $\Tr \left[E^L_i(x) U_i(x)
E^L_i(x+ai) U_i^\dagger(x)\right]$ and leads to the 
simplest improved kinetic Hamiltonian: 
\bea
K_{2}&\!\!\!=\!\!\!& \frac{a^3}{2} \sum_{x,i} \Tr\left[X E^L_i(x)E^L_i(x)
\right. \nn\\
&&\left.+ Y E^L_i(x) U_i(x) E^L_i(x+ai) U_i^\dagger(x)\right].
\eea
To obtain the correct continuum limit and cancel the order $a^2$ errors we
must set $X = 5/6$ and $Y=1/6$. Incorporating the interaction of
arbitrarily distant links leads to an alternative improved kinetic
Hamiltonian:
\bea
 K^{(n)}_{2}&\!\!\!=\!\!\!&
\frac{a^3}{2} \sum_{x,i} \Tr\left[\left( 1 -\frac{1}{6n^2}\right)
E_i^L(x) E_i^L(x) \right. \nn\\
&& \left. \hspace{-1.6cm}+\frac{1}{6n^2}
E_i^L(x)U_{x\rightarrow x+nia}E_i^L(x\!+\!nia)U_{x+nia\rightarrow
  x}\right]\!.
\label{nham}
\eea
An order $a^4$ classically improved kinetic term is given by
\bea
K_{4} &\!\!\!=\!\!\!& \frac{a^3}{2}\sum_{x,i}\Tr\left[
\frac{97}{120}E^L_i(x)E^L_i(x) 
\right. \nn\\
&& \left.\
 + \frac{1}{5}E^L_i(x)U_i(x)E^L_i(x+ai)U^\dagger_i(x) \right. \nn\\
&& \left.\hspace{-1.3cm} 
- \frac{1}{120}
E^L_i(x)U_{\!x\rightarrow x\!+\!2ai}E^L_i(x\!+\!2ai)U_{\!x\!+\!2ai
\rightarrow x}(x)\right]\!\!.  
\label{kimp2}
\eea
Further details are presented in reference~\cite{Carlsson:2001wp}.

\section{Variational calculations}

\begin{figure}[htb]
\vspace{9pt}
\includegraphics[scale=0.6]{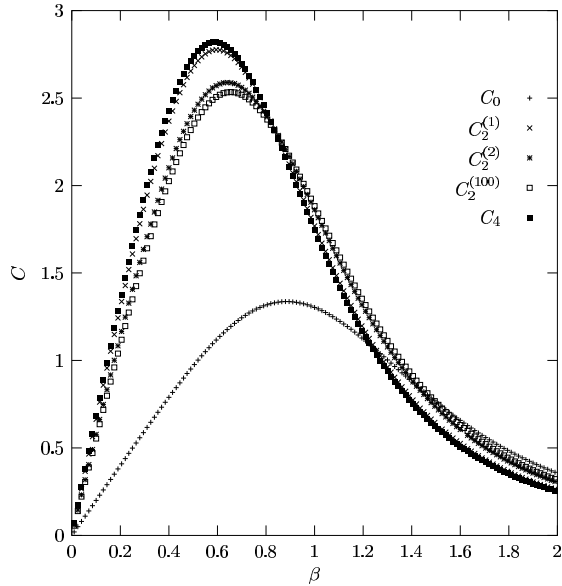}
\caption{SU(2) specific heat in 2+1 dimensions, 
calculated from various Hamiltonians. The
location of the peak indicates the transition region from strong to
weak coupling.}
\label{specificheat}
\end{figure}

We are in the process of performing variational calculations using
improved Hamiltonians. In such calculations we make use of the trial state:
\be
|\phi_0\rangle = \exp \left
(C \Real S_p+ \frac{D}{2}\Real S_r\right)|0\rangle .  
\ee
Here $S_p$($S_r$) is the sum of traced plaquettes (rectangles) on the lattice.
The variational parameters $C$ and $D$ are determined, for a given
coupling, by minimizing the energy density (expectation value of the
improved Hamiltonian per plaquette):
\bea
\epsilon_2
&\!\!\!=\!\!\!& \left\langle\!\!\plaquette\!\!\right\rangle\left(
 \frac{5}{12} 
\frac{N^2-1}{\beta}C - \frac{10\beta}{3N}\right)
\nn\\
&& \hspace{-0.8cm}+\left\langle\!\!\rectangle \!\!\right\rangle\left(
\frac{2}{3}\frac{N^2-1}{\beta} D +\frac{\beta}{3N} \right)
+ 3\beta . \label{edens}
\eea
Here $\beta = N/g^2$. For the case of SU(2) in 2+1 dimensions with $D=0$,
it is possible to express the expectation values analytically in terms
of modified Bessel 
functions ($I_n$), leading to:
\be
\epsilon_2 = \!\left(\frac{5C}{2\beta}\! -\! \frac{10\beta}{3}\right)\!\!
\frac{I_2(4C)}{I_1(4C)} +
\frac{\beta}{3}\!\left(\frac{I_2(4C)}{I_1(4C)}\right)^2 \!\!+3\beta . 
\label{imp}
\ee
This is to be compared with the standard unimproved result of Arisue, Kato and
Fujiwara~\cite{Arisue:1983tt}:
\be
\epsilon_0 = \left(\frac{3C}{\beta} - 2\beta \right)
\frac{I_2(4C)}{I_1(4C)} + 2\beta . \label{unimp}
\ee 
From the results of
Eqs.~(\ref{imp}) and (\ref{unimp}), we calculate the 
respective specific heats, $C_0 =-\partial^2\epsilon_0/\partial^2
\beta $ and $C^{(1)}_2$. Similar results, $C^{(n)}_2$ and $C_4$ are
easily obtained 
from Eqs.~(\ref{nham}) and (\ref{kimp2}) respectively. 
The results are shown in figure~\ref{specificheat}.

The transition region from the strong to weak coupling phase is indicated by
the location of the peak~\cite{Horn:1985ax}. As one would expect,
the improved specific heats have peaks at a larger coupling $g$ than
the unimproved case. It is also clear that
incorporating interactions between distant links in the kinetic term does
not significantly alter the degree of improvement for the case of $D=0$.   
For the case of $D\ne 0$, the expectation values in Eq.~(\ref{edens})
must be calculated numerically. The same is true
for 3+1 dimensional and SU(3) calculations, which are currently in
progress.  

\section{Conclusion}

We have demonstrated the direct improvement of the Kogut-Susskind
Hamiltonian by deriving a number of improved lattice Hamiltonians. Improvement 
has been demonstrated for the simple case of SU(2) in 2+1 dimensions, 
by observing a transition from strong to weak coupling at a larger 
coupling (corresponding to a larger lattice spacing) for various
improved Hamiltonians. Variational calculations of SU(2) and SU(3)
massgaps in 2+1 and 3+1 dimensions using improved Hamiltonians 
are in progress.


\begin{thebibliography}{9}

\bibitem{Kogut:1975ag}
J.~Kogut and L.~Susskind,
Phys.\ Rev.\ D {\bf 11}, 395 (1975).



\bibitem{McKellar:2000zk}
B.~H.~J.~McKellar, C.~R.~Leonard and L.~C.~Hollenberg,
Int.\ J.\ Mod.\ Phys.\ B {\bf 14}, 2023 (2000).

\bibitem{Gregory:2000pm}
E.~B.~Gregory, S.~H.~Guo, H.~Kroger and X.~Q.~Luo,
Phys.\ Rev.\ D {\bf 62}, 054508 (2000).

\bibitem{Lepage:1996jw}
G.~P.~Lepage,
hep-lat/9607076.

\bibitem{Luo:1999dx}
X.~Q.~Luo, S.~H.~Guo, H.~Kroger and D.~Schutte,
Phys.\ Rev.\ D {\bf 59}, 034503 (1999).

\bibitem{Luscher:1985zq}
M.~L\"{u}scher and P.~Weisz,
Phys.\ Lett.\ B {\bf 158}, 250 (1985).
                            %
\bibitem{Carlsson:2001wp}
J.~Carlsson and B.~H.~McKellar,
Phys.\ Rev.\ D {\bf 64}, 094503 (2001).

\bibitem{Arisue:1983tt}
H.~Arisue, M.~Kato and T.~Fujiwara,
Prog.\ Theor.\ Phys.\  {\bf 70}, 229 (1983).

\bibitem{Horn:1985ax}
D.~Horn, M.~Karliner and M.~Weinstein,
Phys.\ Rev.\ D {\bf 31}, 2589 (1985).


\end{thebibliography}
\end{document}